\documentclass[twocolumn]{aastex631}
\usepackage{amsmath}
\hypersetup{
	colorlinks	= true,
	linkcolor	= red,
	urlcolor	= cyan,
	citecolor	= blue
}

\usepackage{amssymb}
\usepackage{xcolor}
\definecolor{mygreen}{rgb}{0.1, 0.7, 0.2}
\definecolor{myorange}{rgb}{1,0.5,0}
\definecolor{myred}{rgb}{1,0,0}

\usepackage{lipsum}
\usepackage{multirow}

\graphicspath{{./images/}}

\newcommand{\exorelr}{\mbox{\textsc{ExoReL$^\Re$}}}

\turnoffediting

\received{May 12$^{th}$, 2023}
\accepted{August 7$^{th}$,  2023}

\submitjournal{AJ}

\shortauthors{Damiano et al.}

\usepackage{fancyhdr}
\begin{document}
	
	% \title{Reflected spectroscopy of small exoplanets III: detection of O$_2$ and O$_3$ in Proterozoic Earth's analogues}
    % \title{Reflected spectroscopy of small exoplanets III: the science case of probing the UV band to measure the biosignature gasses O$_2$ and O$_3$}
    \title{Reflected spectroscopy of small exoplanets III: probing the UV band to measure biosignature gasses}
	
	\correspondingauthor{Mario Damiano}
	\email{mario.damiano@jpl.nasa.gov}
	
	\author[0000-0002-1830-8260]{Mario Damiano}
	\affiliation{Jet Propulsion Laboratory, California Institute of Technology, Pasadena, CA 91109, USA}
	
	\author[0000-0003-2215-8485]{Renyu Hu}
	\affiliation{Jet Propulsion Laboratory, California Institute of Technology, Pasadena, CA 91109, USA}
	\affiliation{Division of Geological and Planetary Sciences, California Institute of Technology, Pasadena, CA 91125, USA}
	
	\author[0000-0003-4205-4800]{Bertrand Mennesson}
	\affiliation{Jet Propulsion Laboratory, California Institute of Technology, Pasadena, CA 91109, USA}
	
	\begin{abstract}
    Direct-imaging observations of terrestrial exoplanets will enable their atmospheric characterization and habitability assessment. Considering the Earth, the key atmospheric signatures for the biosphere is O$_2$ and the photochemical product O$_3$. However, this O$_2$-O$_3$ biosignature is not detectable in the visible wavelengths for most of the time after the emergence of oxygenic photosynthesis life (i.e., the Proterozoic Earth). Here we demonstrate spectroscopic observations in the ultraviolet wavelengths for detecting and characterizing O$_2$ and O$_3$ in Proterozoic Earth-like planets, using \exorelr. %a spectral retrieval framework developed for future exoplanet direct-imaging missions. %This eon represents an intermediate stage between the Archean and Modern Earth, when O$_2$ is much reduced compared with modern Earth while O$_3$ may still be detectable. 
    For an O$_2$ mixing ratio 2 to 3 orders of magnitude less than the present-day Earth, and an O$_3$ mixing ratio of $10^{-7}-10^{-6}$, we find that O$_3$ can be detected and its mixing ratio can be measured precisely (within $~1$ order of magnitude) in the ultraviolet ($0.25-0.4\ \mu$m) in addition to visible-wavelength spectroscopy. With modest spectral resolution ($R=7$) and S/N ($\sim10$) in the ultraviolet, the O$_3$ detection is robust against other potential gases absorbing in the ultraviolet (e.g., H$_2$S and SO$_2$), as well as the short-wavelength cutoff between 0.2 and 0.25 $\mu$m. While the O$_3$ detection does not rely on the near-infrared spectra, extending the wavelength coverage to the near-infrared ($1-1.8\ \mu$m) would provide essential information to interpret the O$_3$ biosignature, including the mixing ratio of H$_2$O, the cloud pressure, as well as the determination of the dominant gas of the atmosphere. %is N$_2$- rather than CO$_2$-dominated. 
    %Our analysis demonstrates that the VIS wavelength band alone is insufficient for characterizing small temperate planets, as many key gases, such as O$_3$, CO$_2$, and CH$_4$, show their main absorption features in the UV or NIR regions of the electromagnetic spectrum. We highlight the critical role of UV observations in accurately detecting O$_2$ and O$_3$, key biosignatures, without the need for NIR observations. Moreover, we confirm the significance of NIR observations for effectively constraining a planet's nature. 
    %We also explored the possibility of the confounding gases H$_2$S and SO$_2$, which also absorb in the UV band, and concluded they do not interfere with the detection of O$_3$ at a spectral resolution of $R=7$. We also explored the UV bandwidth (starting from 0.2 or 0.25 $\mu$m), as reaching far into UV may be an engineering challenge, and concluded that the impact on the detection of O$_2$ and O$_3$ is not significant. 
    %Overall, this work indicates that the capabilities of ultraviolet spectroscopy are essential for characterizing Earth-like exoplanets, and reaffirms the importance of the near-infrared. 
    The ultraviolet and near-infrared capabilities should thus be evaluated as critical components for future missions aiming at imaging and characterizing terrestrial exoplanets, such as the Habitable Worlds Observatory.
	\end{abstract}
	
	
	\section{Introduction} \label{sec:intro}

    High-contrast imaging of exoplanets promises to enable the spectroscopic characterization of temperate and rocky exoplanets in our interstellar neighborhood. Laboratory studies have successfully achieved the necessary level of starlight suppression required for imaging an earth-like planet orbiting a Sun-like star using starshade technology (e.g., \cite{harness2021optical}) and significant progress has been made toward high contrast imaging using coronagraphy (e.g., \cite{trauger2007laboratory,Seo2019}). The Nancy Grace Roman Space Telescope (Roman; \cite{Spergel2015,Akeson2019}) is set to demonstrate high-precision/performance coronagraph technology in space \citep{Mennesson2022}. Future large astrophysics missions combined with starlight suppression technologies could potentially discover small exoplanets in habitable zones around nearby stars and examine their atmospheres across ultraviolet (UV), visible (VIS), and near-infrared (NIR) wavelengths (\cite{Roberge2018,Seager2019,Gaudi2020}; HabEx Final Report\footnote{https://www.jpl.nasa.gov/habex/documents/}; LUVOIR Final Report\footnote{https://asd.gsfc.nasa.gov/luvoir/reports/}). In light of these progresses, the Astro2020 decadal survey recommended the development of a flagship mission to discover Earth-like habitable exoplanets via direct imaging (\cite{national2021pathways}), which is now referred to as the Habitable Worlds Observatory (HWO).

    As such, the characterization of exoplanetary atmospheres through spectroscopy has become a key frontier in the search for potentially habitable worlds. While existing (e.g., JWST) and upcoming telescopes (e.g., Nancy Grace Roman Space Telescope, Roman) have predominantly focused on observations in the visible and near-infrared (NIR) wavelength ranges \citep{ers2022,Feinstein2023,Rustamkulov2023,Alderson2023,Ahrer2023}, the ultraviolet (UV) band offers unique opportunities to probe the presence of crucial atmospheric species, such as O$_2$ and O$_3$, that may serve as potential biosignatures. 
    
    In previous works, the use of reflected light spectroscopy to characterize gas giants, sub-Neptunes, and terrestrial exoplanets has been investigated \citep{Lupu2016,Feng2018,Batalha2019,Damiano2020a,Damiano2020b,carrion2020directly,Damiano2021,Damiano2022,Robinson2023}. These studies demonstrated the effectiveness of using reflected light spectroscopy for the atmospheric characterization of small planets, particularly for characterizing the atmospheres of terrestrial exoplanets beyond modern Earth analogs. Collectively, it has been shown that (1) an optimal definition of Bayesian prior functions leads to a better interpretation of the reflected spectrum by reducing the possibility of degeneracies, (2) The VIS band alone, ranging from approximately 0.4 to 1 $\mu$m, has limitations in fully characterizing terrestrial planets, as it is unable to extract information on CO$_2$ levels, cloud formations, and surface details, (3) the NIR wavelength band ($\sim1-1.8\ \mu$m) is necessary for reliably characterizing terrestrial exoplanet atmospheres from modern Earth-like to Archean Earth-like and CO$_2$-dominated.

    The detection of O$_2$ and O$_3$ in exoplanetary atmospheres is of paramount importance, as their presence can be indicative of biological processes occurring on a planet \citep{meadows2018exoplanet}. The UV band ($\sim0.2-0.4\ \mu$m) provides a promising mean of detecting these species due to their strong absorption in this wavelength range. In the Solar System, the Earth's atmosphere exhibits strong O$_2$ and O$_3$ absorption bands in the UV region \citep[e.g.,][]{Turnbull2006,kaltenegger2007spectral}, which have played a significant role in shaping the planet's ability of a life-bearing world.

    The Earth's atmosphere has experienced several evolutionary phases before reaching its current state. The Archean and Proterozoic eons are two significant periods preceding the present. The Archean Eon, which occurred from around 4 billion to 2.5 billion years ago, is of great importance due to the critical developments in the Earth's crust and the evolution of life during this time. This period saw the creation of the Earth's first stable continental crusts, which contributed to shaping the planet's geological features and set the stage for the development of continents and ocean basins. The atmosphere during this time was characterized by a significant presence of CH$_4$ and CO$_2$ in an N$_2$-dominated scenario (e.g., \cite{catling2020archean}).

    The Proterozoic Eon, spanning from approximately 2.5 billion to 500 million years ago, is also a crucial period in Earth's history. Its significance lies primarily in the critical developments in life's evolution on our planet during this time. These advancements established the groundwork for the emergence of more diverse and complex life forms in subsequent eons. Additionally, the Proterozoic Eon witnessed the Great Oxygenation Event (GOE), which led to a dramatic increase in oxygen levels in the atmosphere. This event had a profound impact on the planet's climate, geology, and habitability for future life forms \citep{Young2009, planavsky2014low}. Nonetheless, the concentrations of O$_2$ and O$_3$ during this period would not have been adequate to display notable absorption signatures in the visible wavelength band, but they would have been sufficient to profoundly affect the UV waveband. Consequently, this scenario is the focus of this study.

    We employ our state-of-the-art Bayesian retrieval method, \exorelr \citep{Damiano2020a,Damiano2021,Damiano2022}, to explore the constraints on O$_2$ and O$_3$ in the atmospheres of terrestrial exoplanets akin to Proterozoic Earth through reflected spectroscopy across UV, VIS, and NIR wavelength bands. Our retrieval framework, which builds upon the methodology presented in the previous works on reflected spectroscopy of small exoplanets \citep{Damiano2021,Damiano2022}, allows us to identify and characterize the dominant atmospheric species without relying on prior assumptions about the background atmosphere. This capability is crucial for accurately assessing the potential habitability of exoplanets with diverse atmospheric compositions.

    To assess the detectability of O$_2$ and O$_3$ in the UV band, we simulate two Proterozoic Earth's analog planetary atmospheric scenarios, which encompass 0.1\% and 1\% of O$_2$ of modern Earth's mixing ratio and the corresponding level of O$_3$ predicted by photochemical models \citep{Reinhard2017}. We explore the sensitivity of our retrieval method to variations in observational parameters such as the wavelength coverage (e.g., the implications of the UV coverage starting at either 0.20 or 0.25 $\mu$m). Additionally, we compare the constraints obtained from observations that include UV with those derived from only visible and NIR bands, to evaluate the added value of including the UV wavelengths in the characterization of exoplanetary atmospheres.

    The paper is organized as follows: Sec.~\ref{sec:model} describes the retrieval algorithm, atmospheric scenarios, and the simulation of reflected light spectra. Sec.~\ref{sec:result} presents the results of our retrievals and the constraints on O$_2$ and O$_3$. In Sec.~\ref{sec:discussion}, we discuss the implications of our findings for the detection of potential biosignatures in exoplanetary atmospheres, as well as the design of future missions aimed at characterizing habitable worlds. Sec.~\ref{sec:conclusion} then summarizes the key conclusions of this study.
	
	\section{Methods} \label{sec:model}
	
	\subsection{Retrieval Setup} \label{sec:retrieval}

    We used \exorelr\ to carry out atmospheric retrievals on synthesized spectra. The details of the algorithm are described in our previous two papers about the atmospheric characterization of small exoplanets, i.e. \cite{Damiano2021, Damiano2022}. The key features of the algorithm are summarized here:

    \begin{itemize}
        \item Implemented centered log-ratio (CLR) of mixing ratios for atmospheric chemical compounds as free parameters;
        \item Developed new set of prior functions \citep[detailed in][]{Damiano2021} to ensure no gas is preferred as the dominant gas a priori;
        \item Included water clouds as type of condensates in the atmosphere;
        \item Used optical properties from \cite{Palmer1974} for cross sections and single-scattering albedo of water droplets;
        \item The volume mixing ratio of water and the cloud density are correlated to ensure physical consistency between water in the gas form and condensation into water clouds;
        \item Cloud parameterization includes cloud top pressure (P$_{top}$), cloud depth (D$_{cld}$), and condensation ratio (CR);
        \item Introduced surface pressure and surface albedo as fixed or free parameters in the model.
    \end{itemize}

    The full list of free parameters used here is reported in Tab.~\ref{tab:retr_par}, along with the ranges and the type of priors for the free parameters.

    \begin{table}[h!]
        \centering
        \begin{tabular}{cccc}
            \hline \hline
            \textbf{Parameter} & \textbf{Symbol} & \textbf{Range} & \textbf{Prior type} \\ \hline
            Surface pressure [Pa] & P$_0$ & 3.0 - 11.0 & log-uniform \\
            Cloud top [Pa] & P$_{top, H_2O}$ & 0.0 - 8.0 & log-uniform\\
            Cloud depth [Pa] & D$_{H_2O}$ & 0.0 - 8.5 & log-uniform \\
            Condensation ratio & CR$_{H_2O}$ & -12.0 - 0.0 & log-uniform\\
            VMR H$_2$O & H$_2$O & -25.0 - 25.0 & CLR$^1$ \\
            VMR CH$_4$ & CH$_4$ & -25.0 - 25.0 & CLR$^1$ \\
            VMR SO$_2$ & SO$_2$ & -25.0 - 25.0 & CLR$^1$ \\
            VMR CO$_2$ & CO$_2$ & -25.0 - 25.0 & CLR$^1$ \\
            VMR O$_2$ & O$_2$ & -25.0 - 25.0 & CLR$^1$ \\
            VMR O$_3$ & O$_3$ & -25.0 - 25.0 & CLR$^1$ \\
            Surface albedo & A$_g$ & 0.0 - 1.0 & linear-uniform\\
            Surface gravity [cgs] & \textit{g} & 1.0 - 5.0 & log-uniform\\ \hline \hline
            \end{tabular}
        \caption{Priors for the free parameters used in the retrievals presented in this work. NOTE - $^1$ \cite{Damiano2021}}
        \label{tab:retr_par}
    \end{table}

    Finally, we used \textsc{MultiNest} \citep{Feroz2009, Buchner2014} as Bayesian algorithm to explore the multi-dimensional free parameter space. We used 1000 live points and a Bayesian tolerance of 0.5 to be able to capture any possible degeneracies.
	
	\subsection{Simulated Atmospheric Scenarios} \label{sec:scenarios}

    In our previous work \citep{Damiano2022}, we presented four different case studies to asses the impact on retrieved parameters by the probed wavelength range for terrestrial exoplanets. Among the four scenarios, we included an Archean Earth-like and a Modern Earth-like atmosphere. To characterize these two scenarios, the use of the UV wavelength band was not required as the Modern Earth shows strong absorption features of both O$_2$ and O$_3$ in the VIS wavelength band, while the Archean Earth does not contain enough O$_2$ and O$_3$ to show significant absorption in either visible or UV wavelengths. In the history of the Earth evolution, the passage from the Archean to Modern environment is characterized by the Proterozoic Eon, when the atmospheric O$_2$ was substantial but lower than the modern Earth's mixing ratio by at least two orders of magnitude \citep{planavsky2014low}. In the Proterozoic Earth-like scenario, while the presence of O$_2$ and O$_3$ is not negligible and a product of the biosphere, they do not have enough atmospheric abundance to produce detectable absorption features in the VIS wavelength band. In this case the UV band can be useful, where small amount of O$_2$ and O$_3$ produce strong signatures \citep{Reinhard2017}. 

\begin{figure}[h!]
      \centering
      \includegraphics[width=0.5\textwidth]{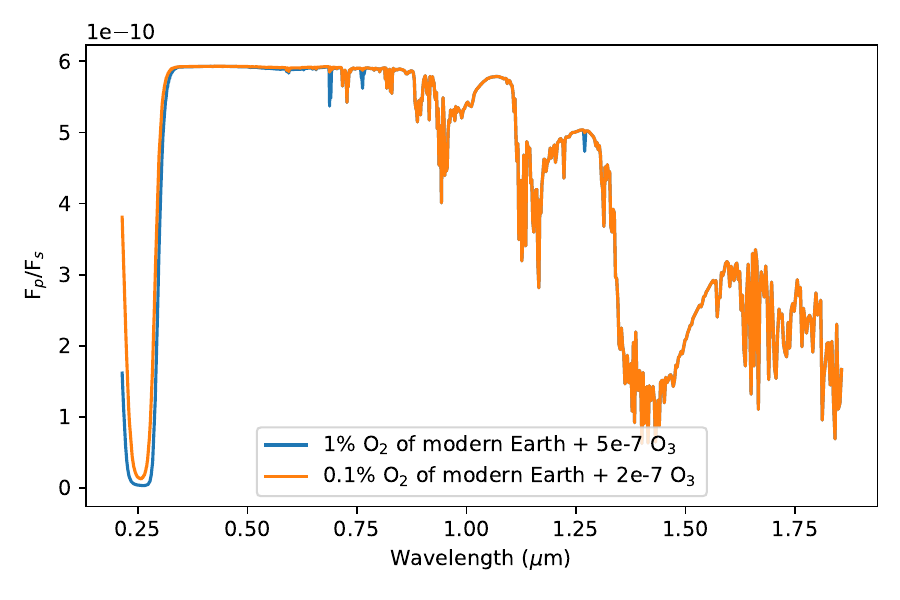}
      \caption{Simulated spectra of the two Proterozoic Earth-like scenario considered in this study.}
      \label{fig:gen_spectra}
    \end{figure}

    Here we simulate a 1M$_{\oplus}$, 1R$_{\oplus}$ planet at 1AU from a Sun-like star. We surround the planet with two different atmospheric Proterozoic-like scenarios to explore the capability of reflected light spectroscopy in different wavelength bands. The first scenario is a Proterozoic Earth-like planet containing 1\% of the O$_2$ concentration of the Modern Earth value, while in the second scenario we dropped the O$_2$ concentration to 0.1\% of Modern Earth concentration value. We modeled the concentration of O$_2$ and O$_3$ to be consistent with each other by considering the results presented in \citep{Reinhard2017}. The true value used to simulate the spectra are reported in Tabs.~\ref{tab:pro_1p} and \ref{tab:pro_01p}, while the generated spectra are shown in Fig.~\ref{fig:gen_spectra}. 
    
\begin{figure}[h!]
      \centering
      \includegraphics[width=0.5\textwidth]{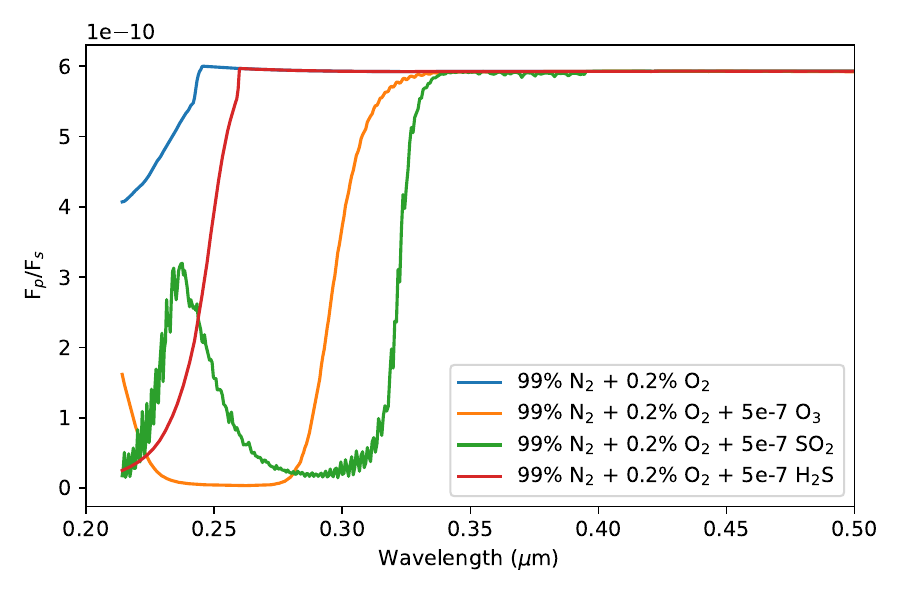}
      \caption{Absorption features of different gases in the UV wavelength band. In particular, O$_3$, SO$_2$, and H$_2$S absorption features are compared.}
      \label{fig:multi_gas}
    \end{figure}

    Finally, in the final reports for HabEx and LUVOIR, concerns have been raised about the potential for degenerate solutions between O$_3$ and other chemical compounds within the UV wavelength range. In particular, SO$_2$ and H$_2$S show strong absorption features close to that of the ozone (Fig.~\ref{fig:multi_gas}). A closer look reveals that the features do not overlap with each other, signaling that degeneracy on the detection of O$_3$ is possible but not likely. To test this assumption, we included SO$_2$ as free parameter of the retrieval to see if we would detect it at a modest spectral resolution ($R=7$) in any of the studied cases.

	In all scenarios, we study the impact of including and excluding the UV and the NIR portion of the reflected light spectra in the retrieval. In this work, we set the ``UV" wavelengths to $0.2/0.25 - 0.4\ \mu$m, the ``VIS'' wavelengths to $0.4-1.0\ \mu$m and the ``NIR'' wavelengths to $1.0-1.8\ \mu$m, and adopt a baseline spectral resolution of R$=$7, R$=$140, and R$=$40 for the UV, VIS, and NIR bands, respectively. These assumptions are approximately consistent with the large mission studies \citep{Roberge2018,Gaudi2020}.

    \begin{figure*}[]
      \centering
      \includegraphics[width=\textwidth]{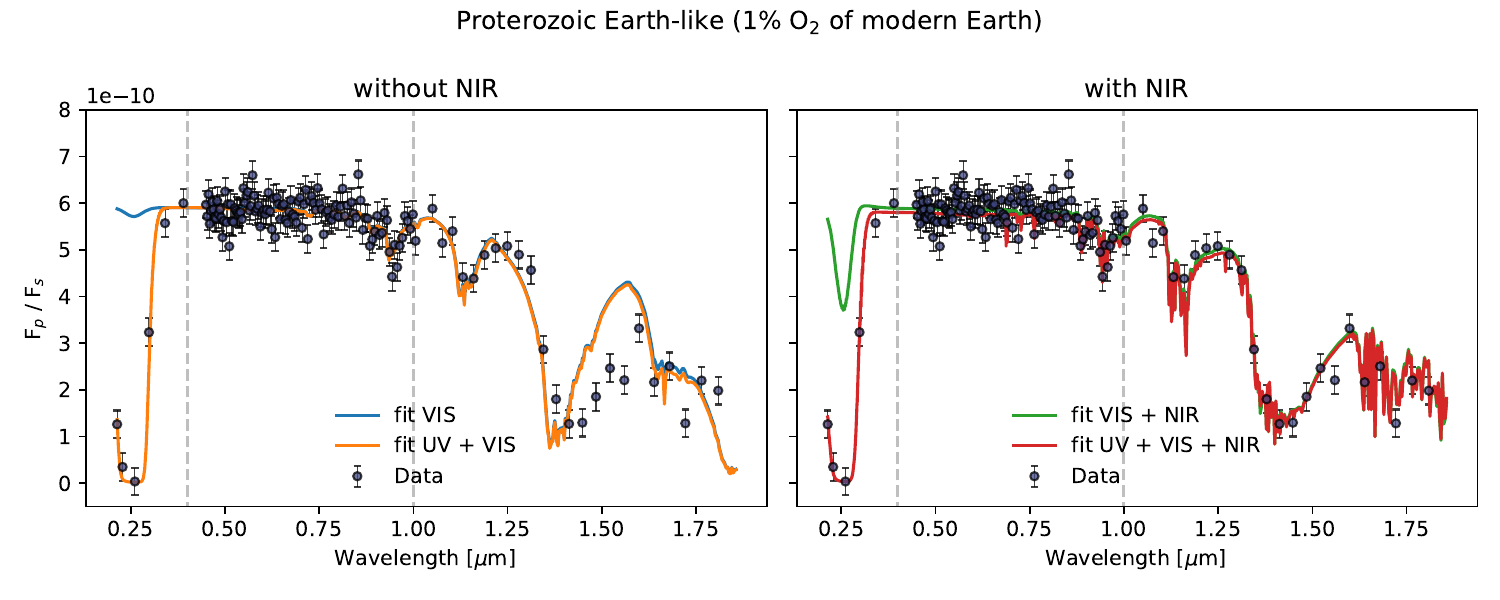}
      \caption{Simulated data and best fit retrievals for the Proterozoic Earth-like scenario with 1\% of Modern Earth O$_2$. VIS/UV/NIR spectrum predicted when fitting: VIS data only (blue curve), UV+ VIS data only (orange curve, same as blue curve, except in the UV), VIS + NIR data only (green curve), UV + VIS + NIR data only (red curve, same as green curve, except in the UV).}
      \label{fig:pro_1p}
    \end{figure*}
	
	Lastly, to simulate the error associated to each of the data points, we considered the maximum value of the spectrum and we divided it by the desired S/N, 20 in this study. This effectively assumes that the noise is dominated by the background and not the planet, likely valid for spectroscopy of terrestrial planets \citep[e.g.,][]{hu2021overview}. In the UV band, this corresponds to an average S/N=9 for the 1\% O$_2$ scenario and S/N=12 for the 0.1\% O$_2$ scenario. We then added a Gaussian deviation to the data to simulate the random realization of the observation.

	\section{Results} \label{sec:result}

    In this section, we present the results of the retrieval process applied to the two scenarios presented in Sec. \ref{sec:scenarios}.

    \subsection{1\% of Modern Earth's O$_2$}

    The retrieval results are reported in Tab.~\ref{tab:pro_1p}, and the posterior distributions are shown in Appendix~\ref{sec:A_1p}. The simulated data and the best-fit spectra are shown in Figure~\ref{fig:pro_1p}.
    
    \begin{deluxetable*}{cccccc}
		\tablecaption{Atmospheric parameters used to simulate the Proterozoic Earth-like scenario (1\% of Modern Earth O$_2$) and the retrieval results. The error bars of the retrieval median results correspond to the 95\% confidence interval (i.e., 2$\sigma$). NOTE - parameters highlighted with the asterisk are derived from the other free parameters. Color code: \textcolor{mygreen}{green} strongly constrained within 3$\sigma$, \textcolor{myorange}{orange} weakly constrained within 3$\sigma$, and \textcolor{myred}{red} not constrained or outside 3$\sigma$. \label{tab:pro_1p}}
		\tablehead{
			\colhead{Parameter} & \colhead{Input} & \colhead{VIS} & \colhead{UV$+$VIS} & \colhead{VIS$+$NIR} & \colhead{UV$+$VIS$+$NIR}}
		\startdata
		$Log(P_{0})$ [Pa] & $5.00$ & \textcolor{myred}{$9.76^{+1.15}_{-1.18}$} & \textcolor{myred}{$9.78^{+1.15}_{-1.25}$} & \textcolor{mygreen}{$5.45^{+0.67}_{-0.52}$} & \textcolor{mygreen}{$5.32^{+0.64}_{-0.41}$} \\
		$Log(P_{top, H_2O})$ [Pa] & $4.85$ & \textcolor{myred}{$3.47^{+3.65}_{-3.19}$} & \textcolor{myred}{$3.48^{+3.74}_{-3.30}$} & \textcolor{myorange}{$5.09^{+0.65}_{-0.52}$} & \textcolor{myorange}{$4.96^{+0.55}_{-0.36}$} \\
		$Log(D_{cld, H_2O})$ [Pa] & $4.30$ & \textcolor{myred}{$3.61^{+3.88}_{-3.34}$} & \textcolor{myred}{$3.79^{+3.93}_{-3.57}$} & \textcolor{mygreen}{$4.65^{+0.69}_{-0.58}$} & \textcolor{mygreen}{$4.52^{+0.56}_{-0.43}$} \\
		$Log(CR_{H_2O})$ & $-3.00$ & \textcolor{myred}{$-5.67^{+5.20}_{-5.83}$} & \textcolor{myred}{$-5.74^{+5.45}_{-5.93}$} & \textcolor{myorange}{$-5.06^{+2.19}_{-4.40}$} & \textcolor{myorange}{$-4.79^{+1.92}_{-4.02}$} \\
		$Log(VMR_{H_2O})$ & $-2.01$ & \textcolor{myred}{$-7.37^{+0.26}_{-0.28}$} & \textcolor{myred}{$-7.35^{+0.26}_{-0.28}$} & \textcolor{mygreen}{$-2,28^{+0.55}_{-0.56}$} & \textcolor{mygreen}{$-2.12^{+0.42}_{-0.44}$} \\
		$Log(VMR_{CH_4})$ & $-4.30$ & \textcolor{myred}{$-7.74^{+0.57}_{-5.77}$} & \textcolor{myred}{$-7.64^{+0.54}_{-5.79}$} & \textcolor{mygreen}{$-4.53^{+0.54}_{-0.73}$} & \textcolor{mygreen}{$-4.37^{+0.39}_{-0.62}$} \\
        $Log(VMR_{SO_2})$ & $-$ & $-$ & \textcolor{myred}{$-11.02^{+2.19}_{-2.80}$} & $-$ & \textcolor{myred}{$-8.71^{+0.83}_{-1.42}$} \\
		$Log(VMR_{CO_2})$ & $-3.40$ & $-$ & $-$ & \textcolor{myorange}{$-5.94^{+2.92}_{-3.88}$} & \textcolor{myorange}{$-5.75^{+2.79}_{-3.84}$} \\
		$Log(VMR_{O_2})$ & $-2.71$ & \textcolor{myred}{$-8.54^{+2.73}_{-5.13}$} & \textcolor{myred}{$-8.44^{+2.73}_{-5.10}$} & \textcolor{myorange}{$-3.84^{+1.77}_{-5.94}$} & \textcolor{mygreen}{$-2.61^{+0.65}_{-6.14}$} \\
		$Log(VMR_{O_3})$ & $-6.30$ & \textcolor{myred}{$-9.61^{+3.48}_{-4.19}$} & \textcolor{mygreen}{$-6.11^{+0.16}_{-0.19}$} & \textcolor{myred}{$-7.73^{+1.75}_{-2.36}$} & \textcolor{mygreen}{$-6.32^{+0.24}_{-0.26}$} \\
		$Log(VMR_{N_2})^*$ & $-0.005$ & \textcolor{mygreen}{$-0.01^{+0.01}_{-0.01}$} & \textcolor{mygreen}{$-0.01^{+0.01}_{-0.01}$} & \textcolor{mygreen}{$-0.01^{+0.01}_{-0.01}$} & \textcolor{mygreen}{$-0.01^{+0.01}_{-0.01}$} \\
		$A_g$ & $0.2$ & \textcolor{myred}{$0.54^{+0.43}_{-0.50}$} & \textcolor{myred}{$0.51^{+0.47}_{-0.48}$} & \textcolor{myorange}{$0.25^{+0.38}_{-0.24}$} & \textcolor{myorange}{$0.22^{+0.39}_{-0.21}$} \\
		$Log(g\ [cgs])$ & $2.99$ & \textcolor{myred}{$3.05^{+0.01}_{-0.01}$} & \textcolor{myred}{$3.05^{+0.01}_{-0.01}$} & \textcolor{mygreen}{$2.99^{+0.02}_{-0.02}$} & \textcolor{mygreen}{$3.00^{+0.02}_{-0.02}$} \\
		$\mu^*$ & $27.93$ & \textcolor{mygreen}{$28.01^{+0.01}_{-0.01}$} & \textcolor{mygreen}{$28.01^{+0.01}_{-0.01}$} & \textcolor{mygreen}{$27.97^{+0.03}_{-0.13}$} & \textcolor{mygreen}{$27.95^{+0.04}_{-0.12}$} \\
		\enddata
	\end{deluxetable*}
    
    We run \exorelr\ on the simulated data while considering different combination of the wavelength band probed. Starting with the VIS band only, we decided not to fit for SO$_2$ and CO$_2$ as the spectral features of these gasses are present in UV and NIR. By inspecting the posterior distribution functions (PDFs) (blue model of Fig.~\ref{fig:post_pro_1p}), the solution on which the retrieval process converged is not compatible with the true values used to synthesize the data in the first place. Even though the code has been able to identify the background gas, i.e. N$_2-$dominated atmosphere, it is not able to constrain the rest of the atmospheric components. The clouds parameters are not constrained at all, showing a flat posterior. This is not surprising as we exposed the weaknesses of considering the VIS band alone in the context of the characterization of the reflected light for small rocky planets in our previous work \citep{Damiano2022}.

    Adding the UV band to the VIS gives information on the O$_2$ and O$_3$ absorption. Looking at the PDFs (orange model of Fig.~\ref{fig:post_pro_1p}), it is possible to see that surface and clouds parameters are still not correctly constrained. However, by inspecting the PDFs of oxygen and ozone, we find that the UV indeed provides the constraint of these two gases, and their PDFs do not show a broad posterior distribution anymore. Finally, we note that fitting SO$_2$, in this case, does not impact whatsoever the constraint on the O$_2$ and O$_3$ as the converged VMR value for SO$_2$ is equal to $-11.02^{+2.19}_{-2.80}$, practically absent.

    Moving on to the VIS plus NIR case (green model of Fig.~\ref{fig:post_pro_1p}), the general nature of the planet is correctly identified: the background gas is correctly found, and the majority of the other gases are constrained except for CO$_2$ and O$_3$ as the VMR of CO$_2$ is too low to show substantial features, and because O$_3$ does not contain absorption bands in the VIS or NIR. Interestingly, O$_2$ is weakly constrained even though its VMR is quite low. This is because O$_2$ shows multiple but small absorption bands in the VIS and NIR wavelength range (see Fig.~\ref{fig:gen_spectra}). The surface pressure  of the planet has been correctly identified as well ass the surface albedo. The cloud parameters are also correctly constrained.

    Finally, adding UV to VIS and NIR observations again measures the mixing ratio of O$_3$. By inspecting the PDFs (red model of Fig.~\ref{fig:post_pro_1p}), we can confirm that the nature of the planet has been correctly retrieved as the totality of the parameters converged to the true values. Also in this case, SO$_2$ does not impact the convergence of the detection of O$_3$.

\begin{figure}[]
      \centering
      \includegraphics[scale=0.48]{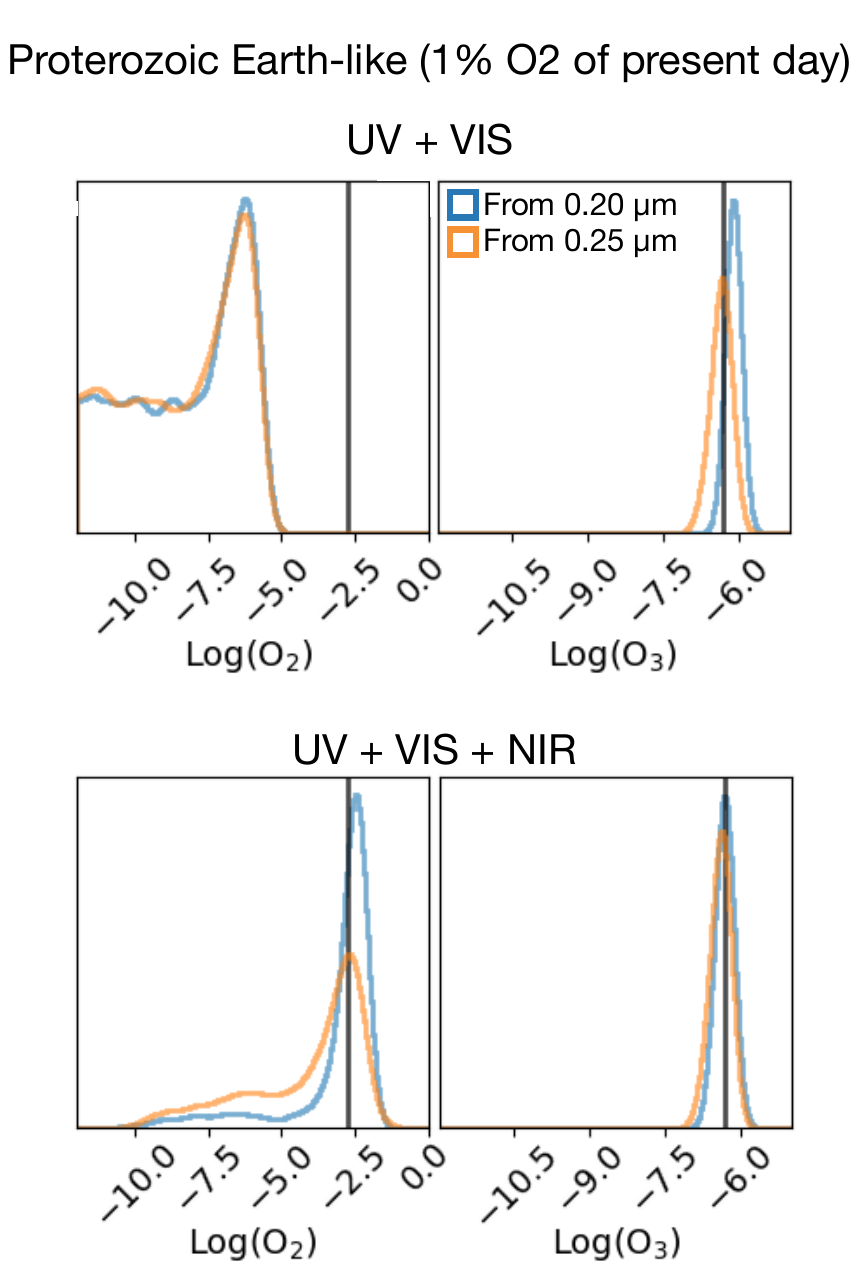}
      \caption{Effect of the lower wavelength cutoff of the UV band on the PDFs of O$_2$ and O$_3$ for the Proterozoic Earth-like with 1\% of Modern Earth's O$_2$. The results of the retrieval does not change appreciably. The constraints are moderately affected as the distributions are broadened. Blue curves correspond to the results shown in Tale 2, with the default 0.2 um UV lower bound.}
      \label{fig:uv_band_1p}
    \end{figure}

    We have also performed a test to reveal effect on the retrieval when considering a different lower limit of the UV band. As it may be difficult to reach 0.2 $\mu$m, we tested the case in which the UV is defined starting from 0.25 $\mu$m. Fig.~\ref{fig:uv_band_1p} shows the effect on the retrieval results of O$_2$ and O$_3$. We find that a UV wavelength range of $0.25-0.4\ \mu$m would also allow measurments of O$_3$, as well as O$_2$ when the NIR coverage is also available. The retrieved PDFs for both gases are broader compared to the UV wavelength range of $0.2-0.4\ \mu$m, and the impact is more evident for O$_2$ as there is a strong absorption feature of O$_2$ between 0.2 and 0.25 $\mu$m (Fig.~\ref{fig:multi_gas}). 

    Finally, we noticed a minor and secondary improvement in the posterior distribution when UV is considered in addition to VIS and NIR bands. In Fig.~\ref{fig:post_pro_1p}, for example, the top pressure of the clouds is better constrained than the VIS+NIR case. However, when UV is considered along with VIS only (VIS vs. UV+VIS), we do not see any improvement. Moreover, we underline that the difference is only noticeable for the 1\% O$_2$ level suggesting a correlation with the detection of O$_2$ in the UV and NIR bands. The improvement vanishes once the UV is cut to 250 nm or for the 0.1\% O$_2$ level.

    \subsection{0.1\% of Modern Earth's O$_2$}

\begin{figure*}[]
      \centering
      \includegraphics[width=\textwidth]{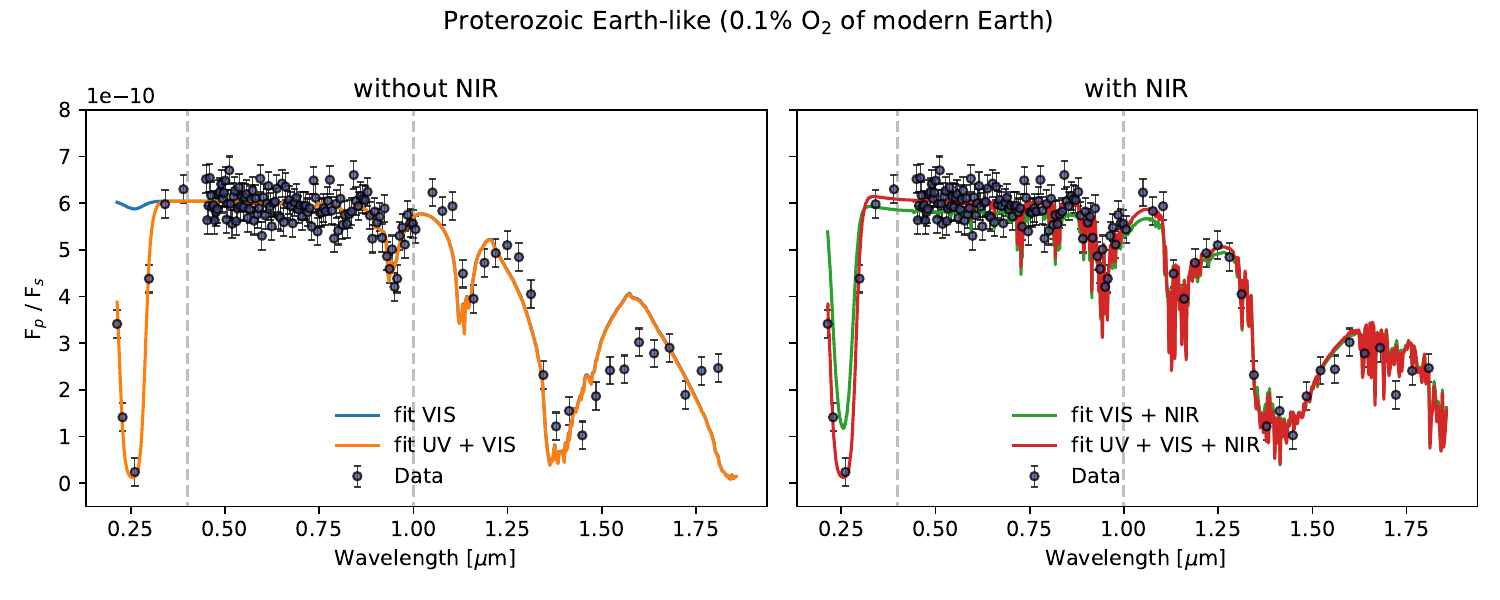}
      \caption{Simulated data and best fit retrievals for the Proterozoic Earth-like scenario with 0.1\% of Modern Earth O$_2$. VIS/UV/NIR spectrum predicted when fitting: VIS data only (blue curve), UV+ VIS data only (orange curve, same as blue curve, except in the UV), VIS + NIR data only (green curve), UV + VIS + NIR data only (red curve, same as green curve, except in the UV).}
      \label{fig:pro_01p}
    \end{figure*}

    The Proterozoic Earth-like scenario with 0.1\% of Modern Earth's O$_2$ is identical to the 1\% case except for the input VMR of O$_2$ and O$_3$. The retrieval results are reported in Tab.~\ref{tab:pro_01p}, and the posterior distributions are shown in Appendix~\ref{sec:A_01p}. The simulated data and the best-fit spectra are shown in Figure~\ref{fig:pro_01p}.

    \begin{deluxetable*}{cccccc}
		\tablecaption{Atmospheric parameters used to simulate the Proterozoic Earth-like scenario (0.1\% of Modern Earth O$_2$) and the retrieval results. The error bars of the retrieval median results correspond to the 95\% confidence interval (i.e., 2$\sigma$). NOTE - parameters highlighted with the asterisk are derived from the other free parameters. Color code: \textcolor{mygreen}{green} strongly constrained within 3$\sigma$, \textcolor{myorange}{orange} weakly constrained within 3$\sigma$, and \textcolor{myred}{red} not constrained or outside 3$\sigma$. \label{tab:pro_01p}}
		\tablehead{
			\colhead{Parameter} & \colhead{Input} & \colhead{VIS} & \colhead{UV$+$VIS} & \colhead{VIS$+$NIR} & \colhead{UV$+$VIS$+$NIR}}
		\startdata
		$Log(P_{0})$ [Pa] & $5.00$ & \textcolor{myred}{$9.67^{+1.25}_{-1.28}$} & \textcolor{myred}{$9.62^{+1.28}_{-1.23}$} & \textcolor{mygreen}{$5.32^{+1.07}_{-0.87}$} & \textcolor{mygreen}{$5.26^{+0.78}_{-0.81}$} \\
		$Log(P_{top, H_2O})$ [Pa] & $4.85$ & \textcolor{myred}{$3.18^{+3.59}_{-2.97}$} & \textcolor{myred}{$3.60^{+3.35}_{-3.39}$} & \textcolor{myorange}{$3.88^{+1.05}_{-2.06}$} & \textcolor{myorange}{$3.84^{+0.84}_{-1.44}$} \\
		$Log(D_{cld, H_2O})$ [Pa] & $4.30$ & \textcolor{myred}{$3.93^{+3.58}_{-3.61}$} & \textcolor{myred}{$3.73^{+3.73}_{-3.51}$} & \textcolor{mygreen}{$4.37^{+0.45}_{-0.61}$} & \textcolor{mygreen}{$4.52^{+0.42}_{-0.40}$} \\
		$Log(CR_{H_2O})$ & $-3.00$ & \textcolor{myred}{$-5.98^{+5.60}_{-5.59}$} & \textcolor{myred}{$-5.65^{+5.33}_{-5.86}$} & \textcolor{myorange}{$-8.40^{+6.29}_{-3.24}$} & \textcolor{myorange}{$-8.64^{+4.93}_{-3.08}$} \\
		$Log(VMR_{H_2O})$ & $-2.01$ & \textcolor{myred}{$-7.00^{+0.23}_{-0.19}$} & \textcolor{myred}{$-6.99^{+0.21}_{-0.18}$} & \textcolor{mygreen}{$-1.51^{+0.98}_{-0.59}$} & \textcolor{mygreen}{$-1.54^{+0.51}_{-0.58}$} \\
		$Log(VMR_{CH_4})$ & $-4.30$ & \textcolor{myred}{$-10.80^{+3.05}_{-4.36}$} & \textcolor{myred}{$-10.84^{+3.07}_{-3.55}$} & \textcolor{mygreen}{$-3.98^{+1.00}_{-0.60}$} & \textcolor{mygreen}{$-4.00^{+0.58}_{-0.54}$} \\
        $Log(VMR_{SO_2})$ & $-$ & $-$ & \textcolor{myred}{$-11.31^{+1.91}_{-3.27}$} & $-$ & \textcolor{myred}{$-8.96^{+0.81}_{-1.91}$} \\
    	$Log(VMR_{CO_2})$ & $-3.40$ & $-$ & $-$ & \textcolor{myred}{$-5.50^{+5.13}_{-3.64}$} & \textcolor{myred}{$-6.96^{+4.94}_{-3.48}$} \\
		$Log(VMR_{O_2})$ & $-3.71$ & \textcolor{myred}{$-10.24^{+3.86}_{-4.35}$} & \textcolor{myred}{$-10.35^{+4.11}_{-4.02}$} & \textcolor{myorange}{$-5.63^{+3.20}_{-3.26}$} & \textcolor{myorange}{$-7.48^{+3.78}_{-3.08}$} \\
		$Log(VMR_{O_3})$ & $-6.70$ & \textcolor{myred}{$-10.22^{+4.34}_{-4.72}$} & \textcolor{mygreen}{$-6.64^{+0.13}_{-0.14}$} & \textcolor{myred}{$-7.12^{+1.52}_{-2.70}$} & \textcolor{mygreen}{$-6.68^{+0.13}_{-0.12}$} \\
		$Log(VMR_{N_2})^*$ & $-0.005$ & \textcolor{mygreen}{$-0.01^{+0.01}_{-0.01}$} & \textcolor{mygreen}{$-0.01^{+0.01}_{-0.01}$} & \textcolor{mygreen}{$-0.01^{+0.01}_{-2.61}$} & \textcolor{mygreen}{$-0.01^{+0.01}_{-0.03}$} \\
		$A_g$ & $0.2$ & \textcolor{myred}{$0.49^{+0.42}_{-0.47}$} & \textcolor{myred}{$0.52^{+0.45}_{-0.48}$} & \textcolor{myorange}{$0.44^{+0.31}_{-0.37}$} & \textcolor{myorange}{$0.28^{+0.38}_{-0.25}$} \\
		$Log(g\ [cgs])$ & $2.99$ & \textcolor{myred}{$3.04^{+0.01}_{-0.01}$} & \textcolor{myred}{$3.04^{+0.01}_{-0.01}$} & \textcolor{mygreen}{$3.00^{+0.01}_{-0.01}$} & \textcolor{mygreen}{$2.99^{+0.1}_{-0.02}$} \\
		$\mu^*$ & $27.92$ & \textcolor{mygreen}{$28.01^{+0.01}_{-0.01}$} & \textcolor{mygreen}{$28.01^{+0.01}_{-0.01}$} & \textcolor{mygreen}{$27.73^{+2.76}_{-0.87}$} & \textcolor{mygreen}{$27.74^{+0.21}_{-0.61}$} \\
		\enddata
	\end{deluxetable*}

    Also in this case, the VIS alone (blue model of Fig.~\ref{fig:post_pro_01p}) will not yield any significant constraints. The cloud and surface parameters are either flat or very broad. The background gas has been identified, but the chemical composition of the amotsphere shows significant biases with respect to the truth.

    Adding UV data points to the visible observations results in the detection and constraints of O$_3$ (orange model of Fig.~\ref{fig:post_pro_01p}), while O$_2$ remains undetected by showing no constraints and a broad distribution that encompass almost the entire range probed. 

    Fitting the NIR band in addiction to the VIS provides constraints on the surface and cloud parameters (green model of Fig.~\ref{fig:post_pro_01p}). The general nature of the planet is identified with constraints on the background gas, i.e. N$_2$, and other minor gases like H$_2$O and CH$_4$. Like the previous scenario (1\% O$_2$), O$_2$ and O$_3$ are not constrained. 

    Including the UV band in addition to VIS and NIR allows for the clear constrain of O$_3$ (red model of Fig. \ref{fig:post_pro_01p}). All the other parameters except for CO$_2$ and O$_2$ are constrained. In this scenario, fitting for the presence of SO$_2$ does not impact the PDFs, highlighting once again that SO$_2$ and O$_3$ are well distinct in the UV band.

    \begin{figure}[]
      \centering
      \includegraphics[scale=0.48]{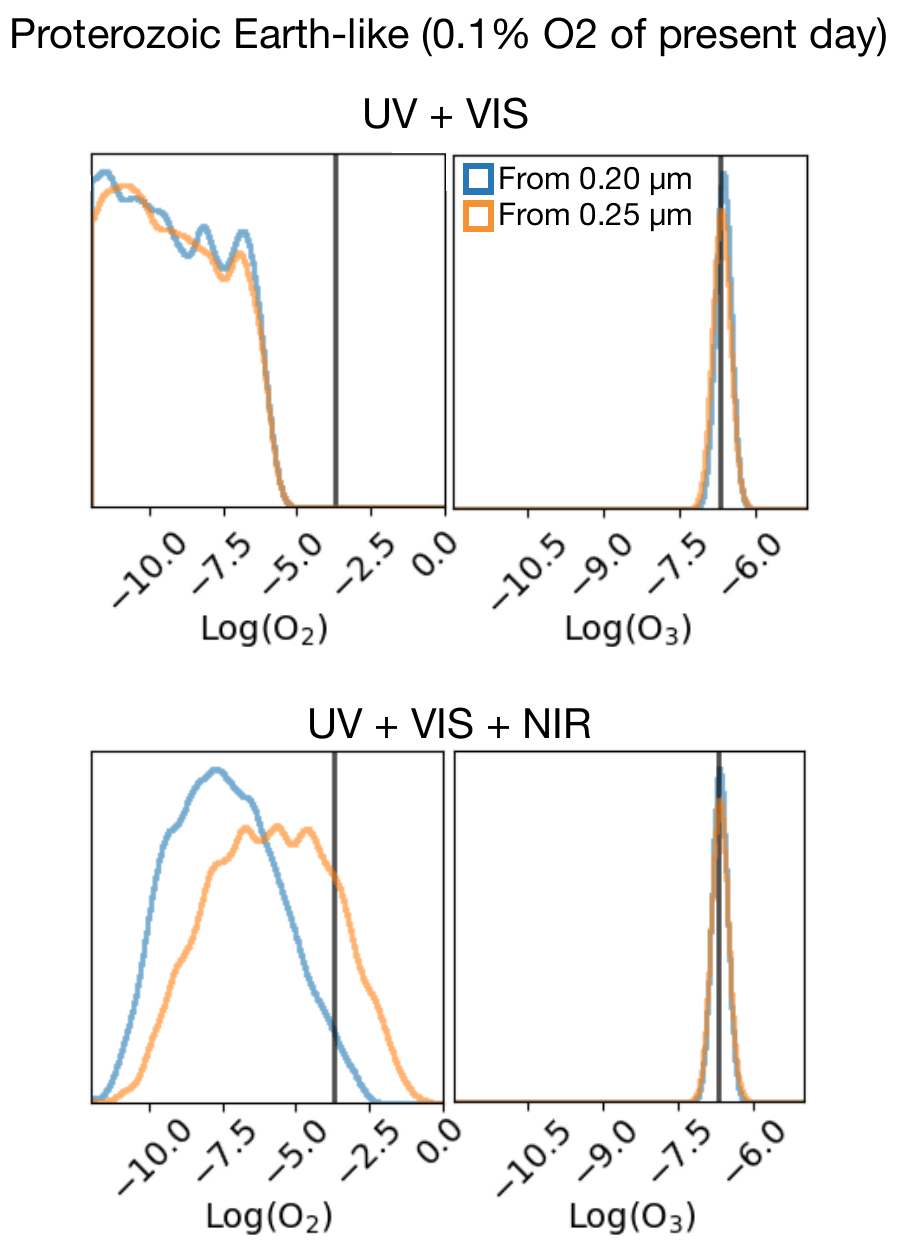}
      \caption{Similar to Fig.~\ref{fig:uv_band_1p} but for the scenario with 0.1\% modern Earth O$_2$.}
      \label{fig:uv_band_01p}
    \end{figure}

    Finally, repeating the retrieval process for a truncated UV band, i.e. starting from 0.25 $\mu$m rather than 0.2 $\mu$m, results in a broadening of the posterior distribution of O$_3$, while O$_2$ remains undetected (see Fig. \ref{fig:uv_band_01p}).
	
	\section{Discussion} \label{sec:discussion}

    The Proterozoic period is placed in between the Archean and the Modern Earth in terms of Earth's atmospheric evolution, and it lasted for $\sim2$ billion years, or $\sim40\%$ of Earth's history. During this period, the biosphere has already been outputting O$_2$ from photosynthesis, but geochemical evidence indicates that atmospheric O$_2$ mixing ratio remained low \citep{planavsky2014low}, preventing remote-sensing detection in the visible wavelengths in similar way to modern-Earth-like planets \citep[e.g.,][]{Feng2018,Damiano2022}. However, due to nonlinear dependency of the photochemical O$_3$ mixing ratio on O$_2$, as well as the strong absorption of O$_3$, spectroscopy in the UV band may detect the O$_2$-O$_3$ biosignature via O$_3$ on a Proterozoic-Earth-like planet \citep{Reinhard2017}. Here we demonstrated, using rigorous spectral retrievals, that the detection of O$_3$ is indeed feasible, and only requires a modest spectral resolution of $R=7$ in a wavelength range of $0.25-0.4\ \mu$m.
    
    This thus gives us a compelling science case for spectroscopic observations in the UV ($0.25-0.4\ \mu$m) for the Habitable World Observatory. 
    Since the atmospheric concentration of O$_2$ and O$_3$ was not sufficient to show significant absorption features in the VIS or NIR wavelength bands for a good part of Earth history, not having the capabilities of starlight suppression and spectroscopy in the UV could mean missing half of Earth-like planets with an oxygenic biosphere. As O$_2$ and O$_3$ remain the leading biosignature gases for terrestrial exoplanets of FGK stars \citep{meadows2018exoplanet}, it is vital to provide sensitivity to O$_3$ at low mixing ratios ($10^{-7}-10^{-6}$, corresponding to the estimated Proterozoic level), which is feasible in the UV.

    %\edit1{The detection of O$_3$ implies the presence of O$_2$, and so, if we could detect O$_2$ directly, the O$_3$ constraints become not as critical. This may be the case for the 1\% O$_2$ levels, where decent (but not good, i.e., broad posterior distribution, -10.0 $\lesssim$ Log(O$_2$) $\lesssim$ -2.0, Fig.~\ref{fig:post_pro_1p}) constraints on the O$_2$ levels can be made with the NIR part of the spectrum along with the VIS part, but without the UV part. 
    %This highlights the importance of the UV channel for scenarios with lower VMR of O$_2$, as the importance of understanding which of those cases existed on Earth, and which of those cases are more likely to exist on exoplanets.}
    
    We also reaffirm the importance of probing the NIR band to successfully constrain the nature of the planet. The addition of the  spectra in $1-1.8\ \mu$m result in the constraints of the surface/cloud properties and, importantly, the mixing ratio of H$_2$O. These pieces of information are essential to establish the hydrological cycle and surface habitability of an exoplanet, upon which the biosignature interpretation would be built. We have retrieved an N$_2$-dominated atmosphere for all scenarios considered in this paper, but for a planet with higher CO$_2$ mixing ratios 
    \citep[e.g., the Archean Earth case in][]{Damiano2022}, adding the NIR band is critical for determining the planet has an N$_2$- rather than CO$_2$-dominated atmosphere. The constraints on H$_2$O and CO$_2$, which are available when the NIR spectroscopy is obtained, also help delineate any geochemical/photochemical false positive scenarios for the O$_2$-O$_3$ biosignature \citep[see a review in][]{meadows2018exoplanet}. In the case of the 1\% O$_2$ levels, decent (but not good, i.e., broad posterior distribution, -10.0 $\lesssim$ Log(O$_2$) $\lesssim$ -2.0, Fig.~\ref{fig:post_pro_1p}) constraints on the O$_2$ levels can be made with the NIR part of the spectrum along with the VIS part, lessening the need for the UV spectrum.

    While not studied here, the UV part of the spectrum can be useful for studying photochemical hazes in planetary atmospheres. Observations in the UV would provide additional information on haze-rich planets such as Archean Earth \citep{arney2016pale}, Titan \citep{trainer2006titan}, Venus \citep{titov2018venushazes}, and Jupiter \citep{artega2022juphazes}). 
    %Although, we did not cover this aspect in this work, it makes another strong case for the UV channel. 
    The detectability of O$_3$ would not be impacted by the presence of hazes as the O$_3$ absorption shows as a sharp edge in the UV while the haze absorption typically creates a slope in the continuum starting from $\sim0.5\ \mu$m. Future studies could assess the information on photochemical hazes (e.g., particle size, absorption coefficient) that could be extracted from the UV and VIS spectra.
    
    %Howerver, Earth itself had an atmosphere very different from the present day for most of its past (e.g. \cite{catling2020archean}), and the scenarios presented in this work demonstrate that having the ability to probe the UV at least up to 0.25 $\mu$m is vital to asses the full nature of the planet. Moreover, the UV yield to the correct detection of O$_2$ and O$_3$ without the needs of observation in the NIR. This makes the UV band an important wavelength window for biosignature search. 

    Taking into account the results of this work and the previous ones \citep{Feng2018,Damiano2022}, it is clear that probing the VIS wavelength band alone is not sufficient for the characterization of terrestrial exoplanets as many key gases, e.g., CO$_2$, CH$_4$, and O$_3$ present their main absorption features in the NIR or UV part of the electromagnetic spectrum. The capabilities in the UV and NIR wavelengths should thus be evaluated as a critical component for the future missions aiming at imaging and characterizing terrestrial exoplanets, such as the Habitable World Observatory.
    
    \section{Conclusions}
	\label{sec:conclusion}

    In this study, we have investigated the potential of ultraviolet (UV), visible (VIS), and near-infrared (NIR) observations to characterize terrestrial exoplanets like the Proterozoic Earth in the context of the direct-imaging spectroscopy in the reflected starlight. This 2-billion-year period of Earth's history is of particular interest, as it represents an intermediate stage between the Archean and Modern Earth, when oxygenic photosynthesis had commenced but the O$_2$ level in the atmosphere remained lower than the modern day.

    Our analysis demonstrates the benefits of UV spectroscopy for detecting and measuring the atmospheric mixing ratios of O$_3$, as well as improving the measurements of O$_2$. Even with a modest spectral resolution of $R=7$ and an average S/N of $\sim10$ for the UV band, the spectra can precisely measure the mixing ratio of O$_3$ to $\sim10^{-7}$, without interference of other UV absorbing gases like SO$_2$. This makes the UV band a crucial wavelength window for the search for O$_2$-O$_3$ biosignatures on terrestrial exoplanets. Furthermore, while the O$_3$ detection does not rely on the NIR spectral coverage, the NIR band would provide essential habitability indicators and contextual information, including the cloud/surface pressure, atmospheric H$_2$O abundances, and dominant atmospheric gases (N$_2$ versus CO$_2$). Overall, it becomes clear that relying solely on the visible (VIS) wavelength band for characterizing small temperate planets can be limiting. This is because many crucial gases, such as O$_3$, CO$_2$, and CH$_4$, display their primary absorption characteristics in the UV (0.25 to 0.4 $\mu$m) or NIR (1.0 to 1.8 $\mu$m) regions of the electromagnetic spectrum.
 
    %Our analysis demonstrates that the VIS wavelength band alone is insufficient for characterizing small temperate planets, as many key gases, such as O$_3$, CO$_2$, and CH$_4$, exhibit their main absorption features in the UV or NIR regions of the electromagnetic spectrum. We have emphasized the importance of UV observations, as they enable accurate detection of O$_2$ and O$_3$ without the need for observations in the NIR, making the UV band a crucial wavelength window for biosignature search. Additionally, we have shown that probing the NIR band is essential for constraining the nature of a planet effectively.

    Based on our findings, future direct-imaging missions aiming at finding and characterizing small rocky exoplanets (e.g., the Habitable World Observatory) should have the capability to probe both the UV and NIR bands, whenever this is feasible from the perspective of the inner working angle. This will significantly enhance our ability to detect and characterize biosignatures, thereby improving our understanding of planetary atmospheres and enhancing our search of life beyond the Solar System.
	
	\section*{Acknowledgments}
	We thank Charles Lawrence, Keith Warfield, Rhonda Morgan, and Shawn Domagal-Goldman for helpful discussion. This work was supported in part by a JPL strategic initiative for developing tools for scientific optimization of missions. This research was carried out at the Jet Propulsion Laboratory, California Institute of Technology, under a contract with the National Aeronautics and Space Administration.
 
    \section*{Software}
    \noindent \exorelr \citep{Damiano2020a,Damiano2021,Damiano2022}, \textsc{Numpy} \citep{Oliphant2015}, \textsc{Scipy} \citep{Virtanen2020}, \textsc{Astropy} \citep{Price2018}, \textsc{scikit-bio} \citep{skbio2020}, \textsc{Matplotlib} \citep{Hunter2007}, \textsc{MultiNest} \citep{Feroz2009,Buchner2014}, \textsc{mpi4py} \citep{Dalcin2021}, and \textsc{corner} \citep{Foreman2016}.

    \newpage
	
	{	\small
		\bibliographystyle{apj}
		\bibliography{bib.bib}
	}

    \appendix
	
	\section{Scenario 1: Proterozoic Earth-like with 1\% of Modern Earth's O$_2$} \label{sec:A_1p}
	
\begin{figure*}[!h]
		\centering
		\includegraphics[scale=0.425]{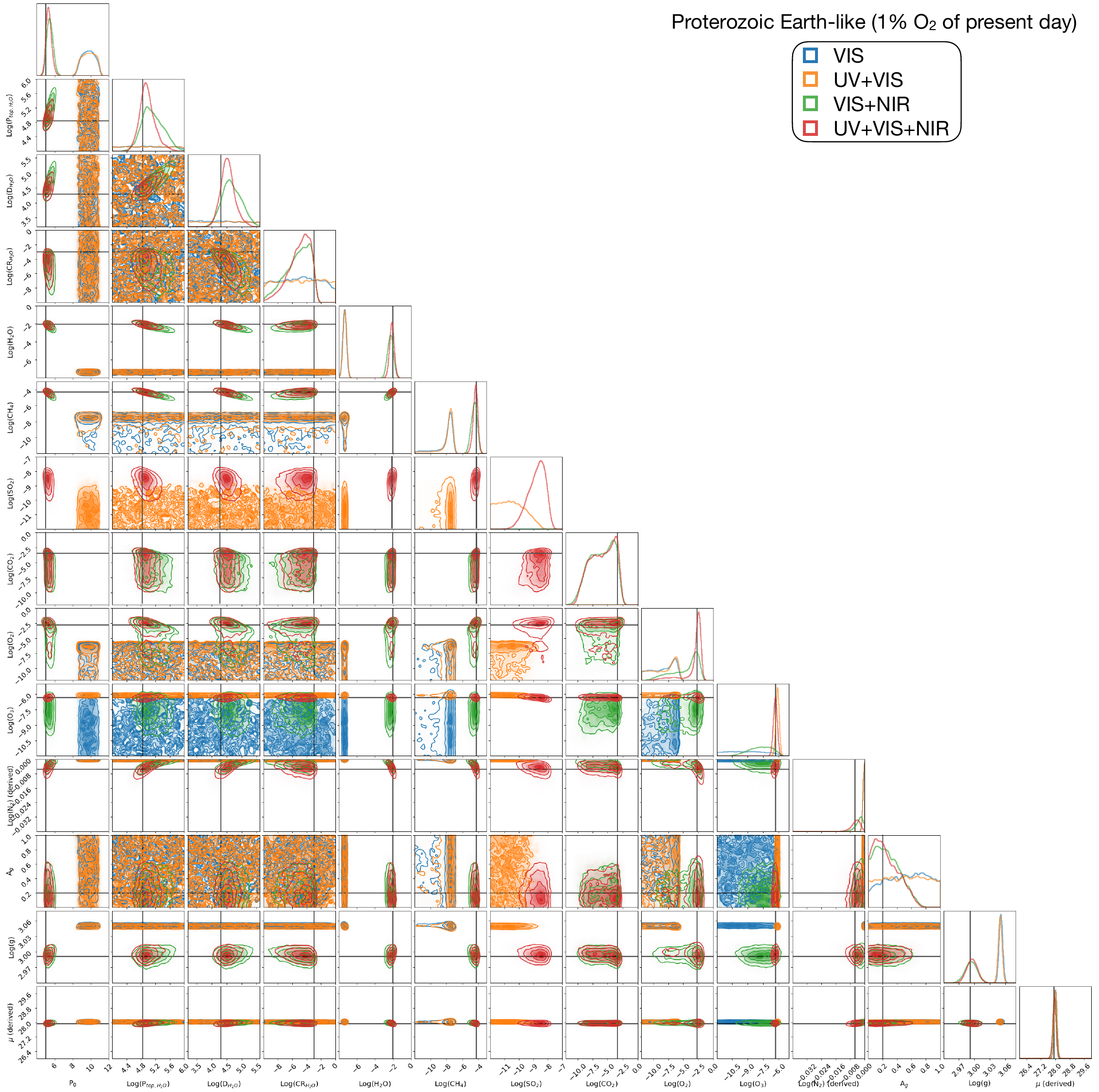}
		\caption{The full posterior distributions (corner plots) of the Proterozoic Earth-like scenario with 1\% of Modern Earth's O$_2$. The different cases explore a combination of different wavelength bands used for the retrieval process. The black lines in the corner plots refer to the true value used to simulate the data. \label{fig:post_pro_1p}}
	\end{figure*}

	Earth's atmosphere has experienced multiple evolutionary stages before reaching the current state. Between the Archean and Modern epochs, another important eon in the Earth's history is called Proterozoic. For this scenario, we simulate the reflected spectrum of an N$_2$-dominated atmosphere (Figs.~\ref{fig:gen_spectra} and \ref{fig:pro_1p}). We also include H$_2$O, CH$_4$, CO$_2$, O$_2$, and O$_3$ as minor absorbing gases. We include a water cloud layer and a surface with albedo of 0.2. We use the synthesized data as input for \exorelr, and Fig.~\ref{fig:post_pro_1p} comprises the resulted posterior distributions of all the wavelength band combinations discussed in Sec.~\ref{sec:scenarios}.
	
	\newpage
	
	\section{Scenario 2: Proterozoic Earth-like with 0.1\% of Modern Earth's O$_2$} \label{sec:A_01p}
	
\begin{figure*}[!h]
		\centering
		\includegraphics[scale=0.425]{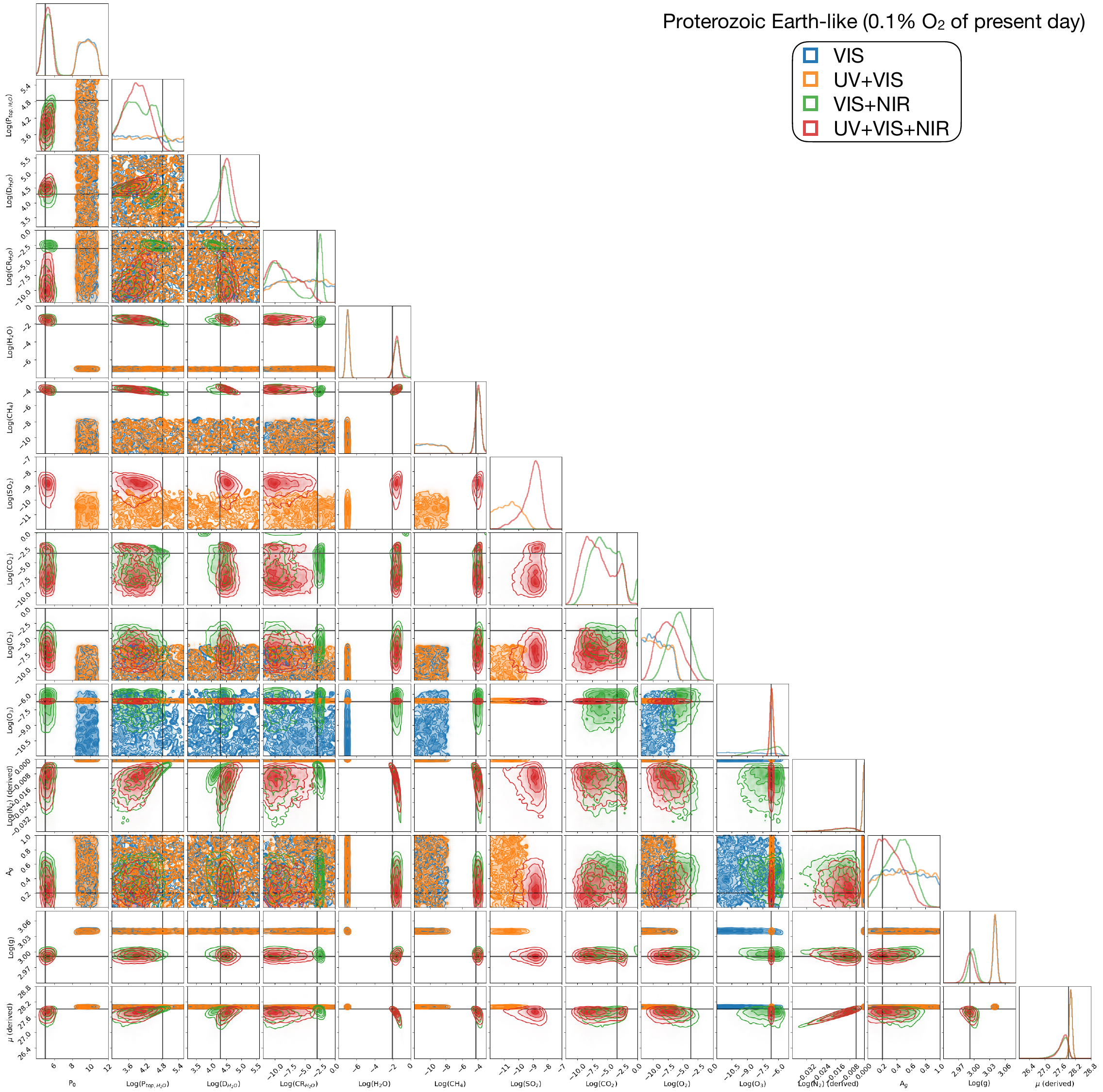}
		\caption{The full posterior distributions (corner plots) of the Proterozoic Earth-like scenario with 0.1\% of Modern Earth's O$_2$. The different cases explore a combination of different wavelength bands used for the retrieval process. The black lines in the corner plots refer to the true value used to simulate the data. \label{fig:post_pro_01p}}
	\end{figure*}

	This scenario is similar to the previous one except for the concentration of O$_2$ and O$_3$. Therefore, we simulate, again, the reflected spectrum of a an N$_2$-dominated atmosphere with H$_2$O, CH$_4$, CO$_2$, O$_2$, and O$_3$ as minor absorbing gases (Fig. \ref{fig:gen_spectra} and \ref{fig:pro_01p}). We still include a water cloud layer and a surface with albedo of 0.2. The synthesized data are used as input for \exorelr. We perform the retrieval with and without the UV and NIR part of the spectrum to explore the benefit of having a larger wavelength band in addition to the optical. The result is shown in Fig.~\ref{fig:post_pro_01p}.
	
\end{document}